\documentclass[12pt,twocolumn,a4paper]{article}

\usepackage{times}
\usepackage{graphicx}
\usepackage{amsmath}

\title{Relativistic positioning systems:\\Perspectives and prospects}
\author{Bartolom\'e Coll\thanks{E-mail: bartolome.coll@uv.es}
\\ \\ \small{Departament d'Astronomia i Astrof\'{\i}sica,} \\ \small{ Universitat de
Val\`encia, 46100 Burjassot, Val\`encia, Spain} }

\begin{document}
\label{paper:Coll}

\maketitle

\begin{abstract}
Relativistic positioning systems are interesting {\em technical objects} for applications around the Earth and in the Solar system. But above all else, they are basic {\em scientific objects} allowing developing relativity from its own concepts. Some past and future features of relativistic positioning systems, with special attention to the developments that they suggest for an {\em epistemic relativity}  (relativistic experimental approach to  physics), are analyzed. This includes {\em relativistic stereometry}, which, together with relativistic positioning systems, allows to introduce the general relativistic notion of (finite) {\em laboratory} (space-time region able to perform experiments of finite size).
\end{abstract}
\section{Introduction}
Relativistic positioning systems were born as {\em scientific objects}.%
\footnote{A scientific object is an object whose knowledge is interesting by itself, independently of its practical utility.}  
But many people consider them as {\em technical objects},%
\footnote{A technical object is an object whose knowledge is interesting for practical applications, 
	   to control our environment.}   
a sort of classical positioning systems directly modeled in relativity. Anyway, their handle is not easy, neither as scientific objects nor as technical ones. This is why to meet all of us together and share methods and ideas is an 	unavoidable step to progress in their development. I want to congratulate the Advanced Concepts Team of the ESA and the Faculty of Mathematics and Physics of Ljubljana for this initiative. This paper is the text of my talk is this meeting.%
\footnote{Workshop {\em  Relativistic Positioning Systems and their Scientific Applications}, held at Brdo in Slovenia from 19th to 21st September 2012.}   

I believe that most people are better interested on relativistic positioning systems as technical objects. But, since their origin as scientific objects, {\em relativistic positioning systems are paradigmatic objects able to become relativity a truly experimental branch of physics}. Relativistic positioning systems are the first component in the construction of {\em relativistic laboratories} of finite size.

It seems evident that to be aware of this role may help us, whatever be our interest, technical or scientific, in solving problems set out by relativistic positioning systems and in finding scientific applications of them.

The purpose of this lecture is twofold:  to present my perspective about some concepts related to relativistic positioning systems, and to prospect the first ingredients for making relativity an experimental approach to physics.
\section{Perspectives}
Why relativistic positioning systems have taken so long to appear? From my personal experience, the answer is very clear: it is due to the obstructions caused by prejudices. We shall begin with a brief account of those physical prejudices that, concerning me, have affected and retarded the natural development of relativistic positioning systems.
\subsection{The genesis of relativistic positioning systems}
The idea of a relativistic positioning system appeared almost simultaneously and in all likelihood independently, in  Bahder \cite{Bahder2001}, Coll \cite{Coll2001} and Rovelli \cite{Rovelli2001}. Every one of these three authors arrived to this idea by very deeply different ways. It is very interesting, and strongly striking, that, in a so short period of time and without apparent precedents, three very similar ideas appeared from three so different ways. 

So different ways limit myself to comment only about {\em my} genesis of the concept.%
\footnote{During many years, the germ of relativistic positioning systems covered a corner of my private garden of thoughts for week-ends and holidays. But every flower sprouted in it, every idea, I showed it to my friends Joan Ferrando, Juan Antonio Morales, Albert Tarantola (\dag  2009) and Jos\'e Mar\' ia Pozo, who watered it carefully. This is the meaning of `{\em my} genesis'.}   
In this genesis, I consider a little number of papers as landmarks or precursors for the ideas presented in  \cite{Coll2001}:
\begin{itemize}
\vspace{-1.mm}

\item[*] {\em Light coordinates in Relativity}  (1985) \cite{Coll1985},

\vspace{-2.5mm}

\item[*] {\em Symmetric frames on Lorentzian spaces} (1991) \cite{CollMora1991},

\vspace{-2.5mm}

\item[*] {\em 199 causal classes of space-time frames} (1992) \cite{CollMora1992},

\vspace{-1mm}
\end{itemize}
They have contributed to the weakening%
\footnote{I would like to say `removal', but recent discussions with colleagues show that it is not the case.}   
of at least one of the following prejudices:
\begin{itemize}
\vspace{-1.mm}

\item[\bf a:]  a physical frame must involve necessarily an a priori definition of space-like synchronization,
\vspace{-2.5mm}

\item[\bf b:]  no frame of four real null vectors exists in relativity,
\vspace{-2.5mm}

\item[\bf c:]  coordinate systems have no physical meaning.
\vspace{-1mm}

\end{itemize}

Prejudice {\bf a} is a mixture of a feeling-based prejudice and an error-based one.  On one hand, it is related to the old feeling, current many decades ago in metrology, that a standard of distance must be matter-based  and not clock-based, and is a remainder of the feeling that an extended instantaneous space is physically meaningful. On the other hand, it comes from a confusion between a physical system and a social or conventional one. It is clear that for our society around the Earth an a priori synchronization  is not absolutely neccesary%
\footnote{The  {\em local Solar time} everywhere on the Earth is an example.}   
but simply convenient. Nevertheless, the absence of a strict symmetry of the gravitational field and of the Earth surface implies the {\em non existence} of an a priori physical synchronization. At the, at present, uncertainty, our conventional synchronization is only possible for our conventional International Atomic Time, not for the proper physical time of every event around the Earth.%
\footnote{Like the Geoid, a physical synchronization on the Earth for a physical time can only be the a posteriori result of continuous careful measures.}$^{\hspace{-0.5mm},}\hspace{-1mm}$
\footnote{Relaxing the space-like condition on a synchronization, i.e. reducing it to the locus of equal time events,  a relativistic positioning system does not defines {\em an} a priori synchronization, but {\em four} equivalent ones.} 

Prejudice {\bf b} is an error-based prejudice, due to  the `saturation' of the concept produced by the abundance of works in the well-known Newman-Penrose formalism of ``null" tetrads \cite{NewmanPenrose1962}. The error consists in applying unconsciously, to {\em any} set of four null vectors, the orthogonality condition imposed by Newman and Penrose to their ``null" tetrads. It is perhaps the weaker one of the above three prejudices, and the easier to dilute, but it has been almost `universal' among relativistic physicists, whatever their renown, up the the last decade. 
 
Prejudice {\bf c} is also an error-based one, due to an incorrect statement of the principle of general covariance. This principle states that {\em the laws of physics are invariant by the choice  of coordinate systems}, and it is an extension of the other one of dimensional invariance, that states their invariance with respect to the particular units used to obtain them. But curiously, during dozens and dozens of years, this statement has slid to the form {\em the laws of physics are independent of the choice  of coordinate systems},%
\footnote{For an object $\omega$, its `independence' of a set $\cal C$ of objects $c$ means that its conception, definition, construction and use may be made in {\em absence} of  $\cal C$, meanwhile its invariance of $\cal C$ means that for its conception or its definition or its construction or its use, objects of $\cal C$ are needed, but that their effect on $\omega$ are {\em independent} of the particular objects $c$, $c'$, etc. of  $\cal C$ taken for its elaboration or use.}  
generating the prejudice in question, meanwhile the similar statement for the dimensional invariance generated a deep research to improve the definition and construction of physical units.%
\footnote{The NIST (National Institute of Standards and Technology), for example, is a good example.} 

On the basis of different combinations of these prejudices, the publication of papers \cite{CollMora1991} and \cite{CollMora1992} was strongly retarded%
\footnote{Almost three years for \cite{CollMora1992}.},     
paper \cite{Coll1985} was forbidden%
\footnote{A hierarchical superior of my research department forbade the submission for publication of an English version of the paper and the continuation of the research on this subject.}   
and the research work on this subject, criticized by many colleagues, was underestimate.

Concerning me, this simple sample of the effects of prejudices already explains in part why relativistic positioning systems have taken so long to appear. But in general the damage that prejudices of referees and colleagues produce  is stronger.
\footnote{Prejudices expend time and money, and demoralize their victims.}  
To help young researches, let me  highlight it :
\begin{quote}
The main obstructions to innovative research are the prejudices. The own prejudices for its conception. Those of the peers for its diffusion.     
\end{quote}

Prejudices do not belong to the past.%
\footnote{There exists no panacea to get rid of them. To remove them is an individual inner process for which, if the intention is neccesary, it is also frequently insufficient. Fortunately, the prejudices cited in the text belong to the class of those that desappear when one is very careful with the analysis of the  conditions under which their assertion is true.}    
The following example concerns a fashion one. 
\subsection{An example of an extended prejudice}
In 2011, the OPERA experiment between the CERN (Geneva) and the LNGS (Gran Sasso) mistakenly reported neutrinos appearing to travel faster than light \cite{Adam2011}.

Some scientists expressed their doubts about this result, like Hawking,%
\footnote{``It is premature to comment on this. Further experiments and clarifications are needed'' said him in  \cite{Collins2011}.}  
or their incredulity, like De Rujula,%
\footnote{``Flabbergasting'', said him in \cite{Overbay2011}.}  
but, as reported in many different media, almost all of them, including the directors of the experiment, believed 
that, if the result were true, relativity theory would be refuted or, at least, deeply damaged. We can thus state, as a general belief among scientists concerned by the subject, that:
\begin{quote}

\vspace{-3mm}
Neutrinos travelling faster than the velocity of light $c$ between CERN and LNGS are inconsistent with standard relativity theory.
\end{quote}

Is this belief correct or is it a prejudice? Let us see it in three steps.

Let us begin remembering what is a local theory. A {\em local theory}  is a theory whose statements and equations are local, i.e. valid in such small space-time%
\footnote{I think that we, physicists, have not yet sufficiently `symbiosed' the concepts of space and time in practise, so that we have not yet merit the moral right to withdraw its hyphen of composed word to the world `space-time'.}   
regions that any physical quantity not mentioned in the statement or not appearing in the equations may be described as constant. Mathematically a physical theory is local if its formulation is infinitesimal (relates physical fields and their space-time variations at every event). 

Now, relativity theory, 
\begin{itemize}
\item[-]  by the concepts used in its construction,
\vspace{-2.5mm}

\item[-]  by the principles on which it is founded,
\vspace{-2.5mm}

\item[-]  by the domain of influence of its equations,
\vspace{-2.5mm}

\item[-]  by the tensor character with which it represents the physical quantities
\vspace{-2.5mm}

\item[-] by the concept itself of space-time that it proposes, 
\vspace{-2.5mm}

\item[-] and because it gives no phenomenological theory for the construction of its  current (energy tensor), but supposes it can be obtained by means of classical balances,
\vspace{-1.5mm}

\end{itemize}
is a local theory in all its constituents.

	Consequently, as all others general statements of the theory, the one that says that {\em the velocity of light $c$ $= 299 792 458$ $m/s$  is a physical limit}, or any other equivalent version, is a local statement. And, being local, the velocity cannot but be an {\em instantaneous velocity}, i.e. measured in a so small time interval that any physical quantity not implied by the concept of velocity is constant. It cannot be, in general, a {\em mean velocity}. The crucial point is that:
\begin{itemize}
\vspace{-1.mm}

\item[*]  In Newtonian theory, where time and space are absolute, if the instantaneous velocity of a particle remains constantly {\em lesser} than a value $v$ during a finite interval, the mean velocity in this interval will also be lesser than $v$. 
\vspace{-2.5mm}

\item[*]  But in relativity, where time and space are different at different events, a particle whose instantaneous velocity  remains constantly lesser than a value $v$, whatever it be,  during a finite interval, may have a mean velocity lesser, {\em equal} or {\em greater} than $v$.
\vspace{-1.5mm}

\end{itemize}
This last fact is easy to see in very simple cases, as it is the one of Fig \ref{FgTwoWays}, representing an accelerated observer submitted to an acceleration $g$ who, at a proper instant $\tau_1$, sends a light signal  to a mirror situated at a proper distance $d$, and receives it at a proper instant $\tau_2$. Because accelerated clocks slow down with respect to inertial ones, and that the distance $d$ to the mirror is greater than that of the inertial observer that cross at $\tau_1$ and $\tau_2$, the mean velocity, $v_m \equiv d / \Delta\tau$, of the light with respect to the accelerated observer can not but be greater than $c$. The precise amount is given by:
\begin{equation*}\label{meanvelocity} 
v_m = c \frac{\frac{g d}{c^2}}{\ln (\frac{g d}{c^2} + 1)} \approx \frac{c}{1 - \frac{1}{2}\frac{gd}{c^2}}.
\end{equation*}
\begin{figure}[h] 
\centering 
{\includegraphics[width=0.40\textwidth,height=0.75\columnwidth] 
{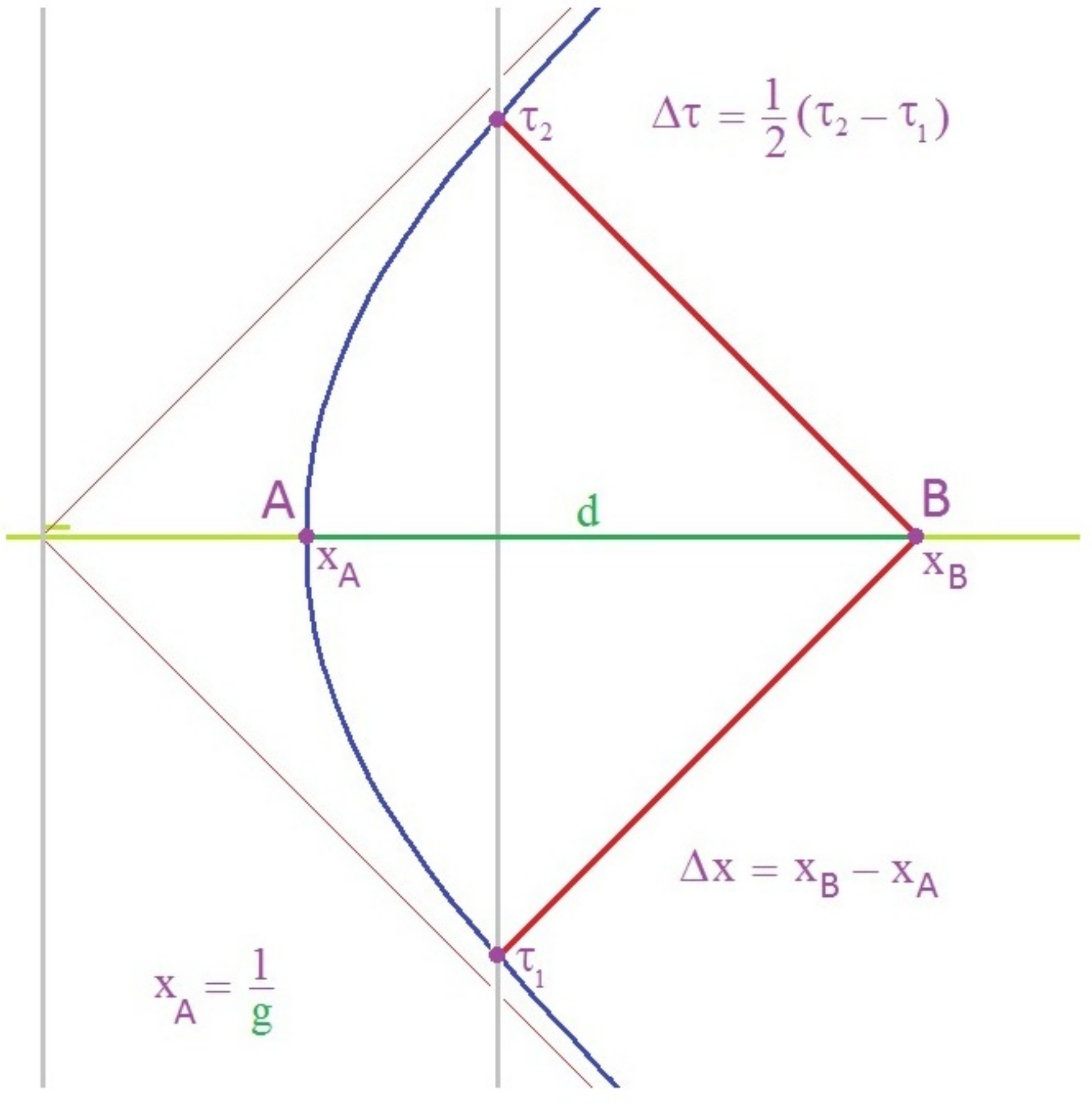}}
\caption{\small An observer A with acceleration $g$ sends at $\tau_1$ a signal to a mirror B situated at a distance $d$, and receives it at $\tau_2$. For him, the mean velocity of light is greater than $c$.}
\label{FgTwoWays}
\end{figure}  

Note that this is a two-way measure that involves only {\em one} clock, {\em one} space and, in it, {\em one} distance. A one-way measure involves in general, not only {\em two} clocks, {\em two} spaces and {\em two} distances,%
\footnote{Take into account that, in relativity, between two observers $A$ and $B$, the spatial distance from $A$ to $B$ is the same than that from $B$ to $A$ only if they are locally near, otherwise they are generically different.} 
but above all, it involves their precise correspondence between them.%
\footnote{In the case of the one-way OPERA experiment, the symmetries (staticity and spherical symmetry) allow to reduce this correspondence to a synchronization between the two clocks.} 

In the OPERA experiment, the mean velocity of  neutrinos, traveling a region of a non constant gravitational field, is measured between two events, the CERN and the LNGS, of different gravitational acceleration (different clock rates). This situation has been modeled in a relativistic gravitational space-time by All\`es \cite{Alles2012} and L\"ust and Petropoulos \cite{Lust2012}, and in both cases they have found mean velocities greater than $c$ for the OPERA configuration, although of many orders of magnitude lesser than the experimental value.

The error of the Opera experiment has not been, as generally believed, to obtain a velocity greater than $c$ for neutrinos,  as relativity foresee,  but the simple quantitative one of obtaining an inappropriate numerical value. 

Let me emphasize this point:
\begin{quote} 
{\em In relativity, mean velocities of particles may be lesser, equal or greater than the instantaneous velocity of light.}
\end{quote}
\subsection{Relativistic and classical positioning systems}
Many people consider relativistic positioning sytems as Newtonian or classical positioning systems directly worked  out with relativity. This is not correct. Although intimately related, they are very different objects. Let us see it.

The aim of a relativistic positioning system is:
\begin{itemize}
	\item[-] to allow any user to know its location in a well defined four-dimensional physical coordinate system,
	\item[-] to provide the user with its proper time and proper distance (space-time metric),
	\item[-] to characterize its space-time trajectory dynamically (proper acceleration) and/or gravitationally (gravimetry).  
\end{itemize}

A relativistic positioning system around the Earth, or RGNSS (Relativistic Global Navigation Satellite System),  wants thus  to characterize the physics of the space-time region between the constellation of satellites and the Earth surface. Fig 2 represents the gravitational field of this extension in an intuitive form.
\begin{figure}[h] 
\centering 
{\includegraphics[width=0.40\textwidth,height=0.50\columnwidth] 
{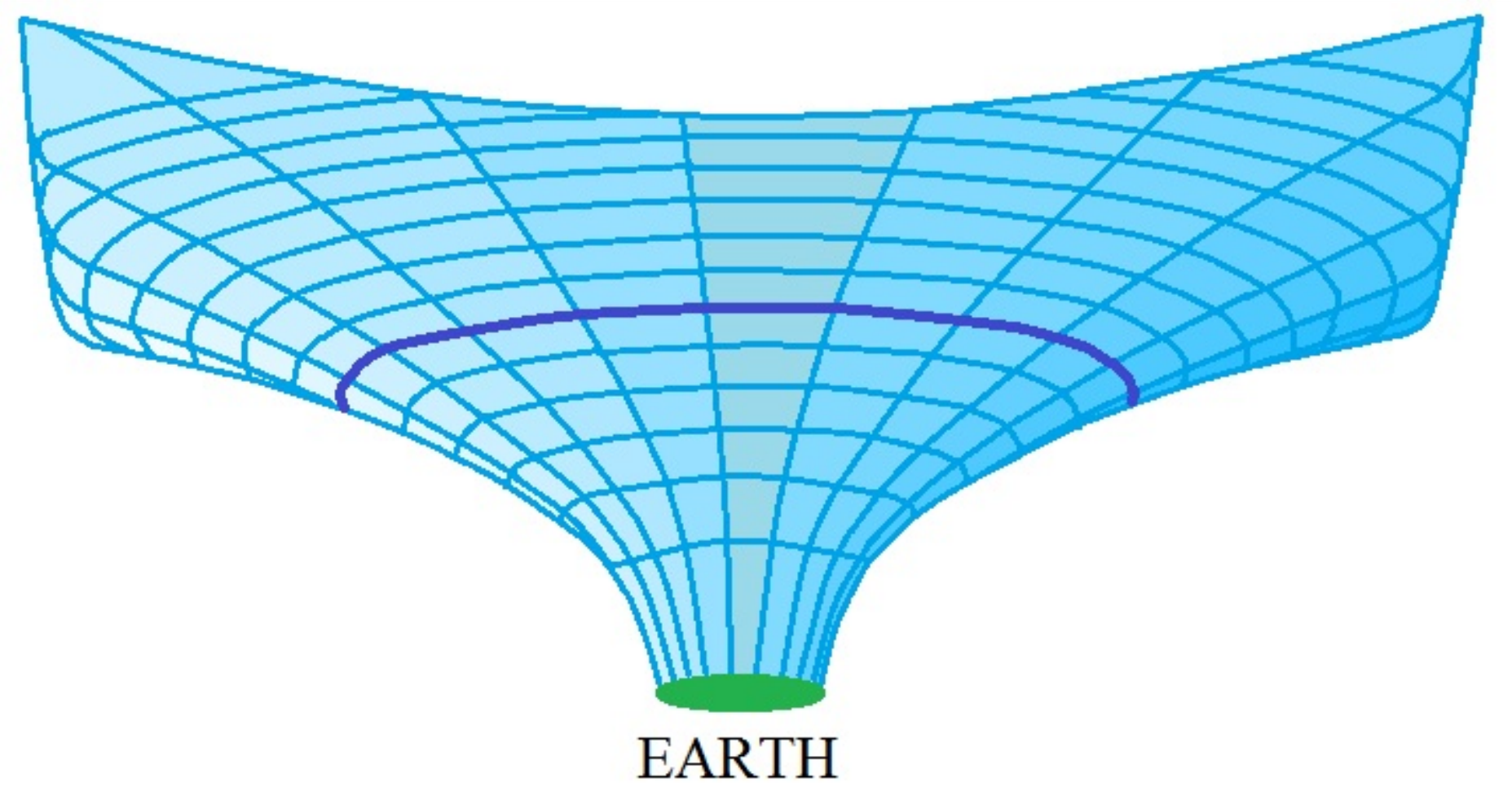}}
\caption{\small The aim of a RGNSS  is to locate users and to provide them with their proper time and proper units of distance  so as to characterize physically the region between the Earth and the satellite constellation.}
\label{FgTwoWays}
\end{figure}  

On the other hand, the aim of a classical positioning system is:
\begin{itemize}
	\item[-] to allow any user to know its position with respect to a specific chart of the Earth surface, and its time with respect to a time scale based on the International Atomic Time (TAI).
\end{itemize}
In the positioning systems around the Earth, or GNSS (Global Navigation Satellite Systems), the specific charts of the Earth surface in use are the World Geodetic System (WGS84) or the International Terrestrial Reference Frame (ITRF), differing by less than ten centimeters in their last determinations. The TAI scale, partially physical, social and political, is a weighted average from many national laboratories clocks,  represents a sort of mean proper time on a mean sea surface level and is much more stable that any individual clock. Moreover, its extension all over the space-time region between the Earth surface and the satellite constellation has undoubtedly many practical and social advantages. 

But this extension at any altitude of the TAI scale and of the Earth surface chart is tantamount to a Newtonization of the space-time region between the Earth surface and the satellite constellation. Fig 3 represents intuitively this situation.
\begin{figure}[h] 
\centering 
{\includegraphics[width=0.40\textwidth,height=0.50\columnwidth] 
{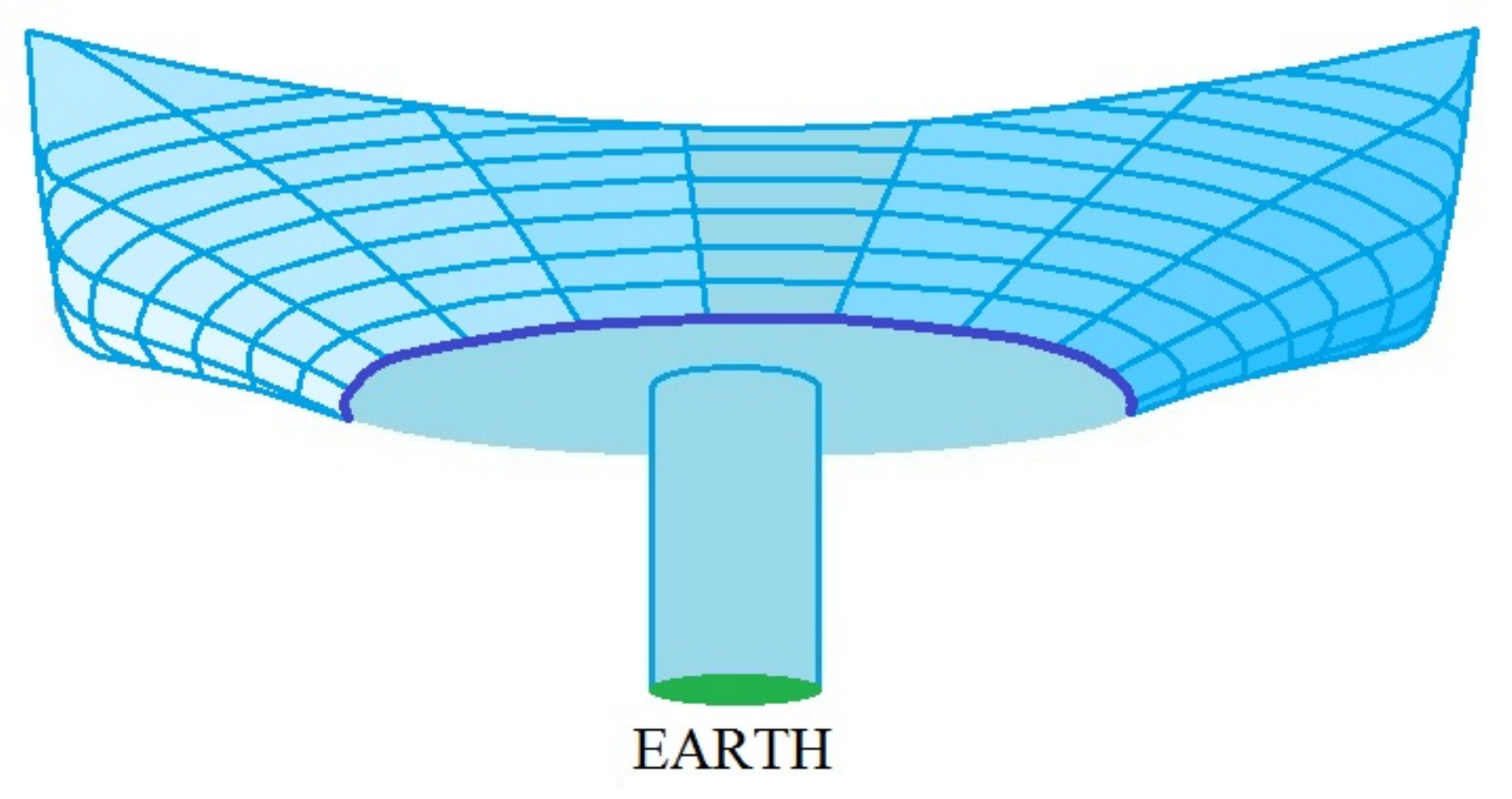}}
\caption{\small The extension of the Earth surface chart and of the TAI to all the space up to the satellite constellation, constitute a Newtonization of this region.}
\label{FgTwoWays}
\end{figure}  

Note that other Newtonizations are possible, as for example the one obtained by the extension to all the region of an averaged time at the satellite constellation level.

Although for a smooth running of both systems, RGNSS and GNSS, the {\em same} information is needed, this information is obtained, interpreted and used very differently. Thus, the physical timing of proper clocks of the satellites constitutes the basic data for relativistic positioning systems, meanwhile this timing is used in classical positioning systems to construct the TAI timing at the satellite level, a pure conventional timing at this level. In this sense, the relativistic corrections for the GPS (for example, in Ashby \cite{Ashby2003}) are used ``sustractively'' on the physical clocks in motion not to improve their precision but, on the contrary,  to better simulate their Newtonian (absolute) behaviour.

This situation may be described in short by saying that relativistic positioning systems are {\em physical} systems meanwhile classical positioning systems are  {\em conventional} ones. To improve both of them, I believe better to first improve the physical systems  without regard to the conventional ones, and then to use these results to improve the conventional systems. Anyway, we must be conscientious with what we are doing: using relativity to better `Newtonize' a GNSS or using relativity to construct a positioning system.                               
\subsection{The main relativistic positioning systems}
Let us remember some known concepts about location systems. A {\em location system} is a physical realization of a coordinate system (for a epistemic definition without reference to mathematics see, for example \cite{Coll2006}).

Two important classes of location systems are the {\em reference systems} and the {\em positioning systems}. The goal of reference systems is to situate the events of a domain with respect to a given observer (generally located at the origin), meanwhile the goal of positioning systems is to indicate its own position to every event of the domain. In Newtonian theory, as far as the velocity of light is supposed infinite, both goals are exchangeable in a sole location system. But in relativity this is no longer possible and it is impossible to construct a positioning system starting from a reference system, but one can always (and very easily) construct a reference system starting from a positioning system. 
\subsubsection{Relativistic positioning systems}
Positioning systems are  {\em immediate}, what means that every event of their domain may know its proper coordinates without delay (in fact, it is this property that defines them). Here they are also supposed {\em generic} and (gravity) {\em free},  guaranteeing their existence in any generic space-time and their construction without the previous knowledge of the gravitational field respectively. It follows from these and the above properties that, whenever possible, it is a positioning system, and not a reference system, that has the most interest to be constructed. From now on, we consider only relativistic positioning systems. Denote by $\cal P$ the set of all of them.
\subsubsection{Auto-locating positioning systems}
An important subclass of positioning systems are the {\em auto-locating positioning systems}, that broadcast their proper time but also the proper time that they receive from their neighbouring satellites.   
Let $\tau^{IJ}$, $I\neq J$, be the proper time of the satellite’s clock $J$ received by the satellite $I$  at its proper time instant $\tau^I$. Then, the sixteen data $\{\tau^I, \tau^{IJ}\}$ received by an observer contains, of course, the emission coordinates $\tau^I$, $(I$ $ = 1, \cdots , 4)$ of this observer but also the coordinates $\{\tau^I,\tau^{IJ}\}$ of every satellite $I$ in the emission coordinate system $\{\tau^I\}$.%
\footnote{The world-lines of the satellites do not belong to the emission coordinate domain of the positioning system, but to its border. Although they are not differentiable along the world-lines, the emission coordinates are well defined on them.}   
Denoting by $\cal L$  the set of all of them, we have  $\cal P$ $\supset \cal L$. 
\subsubsection{Autonomous positioning systems}
Auto-locating systems allow any user to draw univocally the world-lines of the satellites in the emission coordinate system that they broadcast.  But the user still does not know how to draw these world-lines in the space time in which he is living.

For a user to be able to do this, the coordinate data $\{\tau^I, \tau^{IJ}\}$ broadcast by the auto-locating system has to be completed with:
\begin{itemize}
	\item[*] dynamical data of the satellites (acceleration, gradiometry),
	\item[*] observational data from them (e.g. position of reference quasars or pulsars) and
	\item[*] gravitational knowledge of the coordinate region (theoretical, experimental or mixed).
\end{itemize}
The set of this information is called the {\em autonomous data}. Auto-locating systems broadcasting autonomous data are called {\em autonomous positioning systems}. Denoting by $\cal A$  the set of all of them, we have  $\cal P$ $\supset \cal L$ $\supset \cal A$. 

Generic positioning systems, those in the difference $\cal P$ $\setminus \cal L$, have the interest of  have shown that relativistic positioning systems generic, free and immediate exist and, above all, have the advantage of being easier to study than the auto-locating systems constituting $\cal L$. But we have seen on one hand that, whenever possible, there are them, and not reference systems, that have the most interest to be constructed and, on the other hand, we have seen that because the absence of autonomous data they need to be referred to a reference system. That is to say: {\em generic positioning systems are incoherently incomplete or insufficient.} Generic auto-locating systems, those in $\cal L$ $\setminus \cal A$, also inherit the above incoherent incompleteness. 

Thus, non-autonomous positioning systems, those in $\cal P$ $\setminus \cal A$, appear as intermediate hibrids between relativistic reference systems and autonomous positioning systems. 

Autonomous positioning systems are the best location systems. They are the challenge. They were proposed in \cite{Coll2001}.
\section{Prospects}
I do not approve the way relativity has been developed during its century of existence.

Relativity is a physical theory of the gravitational field, but it  is also a physical theory of the space-time.  And it is well stablished that the relativistic descriptions of  both objects, gravitational field and space-time,  improve their corresponding homologues in Newtonian theory. For this reason, I think that:
\begin{itemize}
	\item[*] as a dynamic physical theory, relativity must provide more experiments than simple experimental verifications from time to time, as it is the practice today,
	\item[*] as an improved theory of the space-time, {\em any} physical experiment, whatever it be, {\em ought} to be qualitatively {\em described in the framework of relativity}, regardless of its quantitative evaluation, for which in many cases Newtonian calculations could suffice,
	\item[*] as an improved theory of the gravitational field, relativity {\em ought} to propose experiences and methods 	of measurement of general gravitational fields (four-dimensional metric), which, up to now, are 	conspicuous by their absence.
\end{itemize}
In short, {\em relativity needs to develop a proper experimental approach to the physical world}.
And I believe that we already have the conceptual basic ingredients for this development.

 Now, for this purpose, we need to make more precise the idea of a relativistic experimental approach.
\subsection{Epistemic relativity}
In relativity, a good deal of  scientific works analyze physical and geometrical properties of the space-time, but
\begin{itemize}
	\item dont integrate the physicist as a part of it,
\end{itemize}
and
\begin{itemize}
	\item forget implicitly that:
\begin{itemize}
\item[-] information is energy, 
\item[-] neither the density of energy, nor its velocity
of propagation can be infinite in relativity.
\end{itemize}
\end{itemize}
Many of these properties of the space-time may be analyzed by a {\em geometer} on his desk,  but to be known by an {\em experimental physicist} would require the qualities of an {\em omniscient god !} 

For these reasons, we characterize these scietific works as belonging to {\em ontic relativity}.%
\footnote{From Greek `{\em ontos}', `being', with the meaning of `what it is' as opposed to `how it is seen'.}

Of course, ontic relativity is absolutely {\em necessary} for physics. The conceptual evaluation of many physical situations in order to be able to conceive physical experiments belongs to ontic relativity. But ontic relativity is also manifestly {\em insuficient} for a relativistic experimental approach to physics.

At the opposite side, the works in relativity that:
\begin{itemize}
	\item  integrate the physicist as an element of the problem considered,
	\item concern physical quantities that the physicist can know or measure and
	\item take into account explicitly what information, when and where, the physicist is able to know, 
\end{itemize}
will be considered as characterizing {\em epistemic relativity}.%
\footnote{From Greek `{\em episteme}', `knowledge', with the meaning of `how we obtain it´.}  

 The main objective of epistemic relativity is to provide the physicist with the knowledge  and protocols necessary to make relativistic gravimetry in its (a priori unknown) space-time environment. 

This is the first and unavoidable step to develop experimental relativity as the natural scientific approach to our physical world.
\subsection{Relativistic stereometry}
We know that in the space-time, the adequacy of a mathematical model and the physical system that it describes needs of a univocal correspondence between them. Thus, because, in the differentiable manifold of the mathematical model, points are identified by their coordinates, we need to know how to construct a {\em location system}, that is to say, how to, correspondly, label the events of the physical space-time. But we know also that the best location systems (those which are generic, free and immediate) 
are the relativistic positioning systems. Consequently, it becomes evident that relativistic positioning systems are the first ingredient of epistemic relativity.

	What other else do we generically need in epistemic relativity? 
\subsubsection{A finite laboratory}
In fact, what we need is to be able to consider the space-time region of physical interest 
as a laboratory. The question is then: what is a laboratory (of finite dimension) in relativity? 
	
A simple reflexion shows that, in fact, and regardless of the specificity of its measurement devices, any laboratory, has to provide us with:
\begin{itemize}
	\item[*] a precise location of the significant parts of the physical system in question, and 
	\item[*] a precise description of its pertinent intrinsic physical properties. 
\end{itemize}

	Similarly to the precise location, which is obtained with a system of four clocks (relativistic positioning system), the precise description of the intrinsic properties of a system has to be obtained with a system of four (relativistic) observers. Such a system of four observers is called a {\em stereometric system}. Thus,
\begin{quote} 
 A finite {\em laboratory} in relativity is a space-time region endowed with 
\begin{itemize}
\item[*]  a relativistic positioning system and 
\item[*] a relativistic stereometric system. 
\end{itemize}
\end{quote}
\subsubsection{Relativistic stereometric systems}
In physics, the word `observer' is rather polysemic. To what notion of observer are attached the  relativistic stereometric systems?

Here an {\em observer} is a $4\pi$-wide hypergon {\em eye}%
\footnote{Also called $4\pi$-steradian fish eye.}  
able to record and to analyze its input. 
It is a {\em local} device,%
\footnote{Physically a device is {\em local} if it takes up such a small space-time region that all physical fields in it may be considered as constants. Mathematically it means that it needs only of an infinitesimal region around a space-time event to be defined.}  
defined at every space-time event by its unit velocity, that projects the past light cone of the event on its {\em celestial sphere}.%
\footnote{The celestial sphere of an observer at a space-time event is the quotient of its three-dimensional space by the set of all the past null directions converging at this event.}   

Although they are not very abondant,  there exist interesting papers on relativistic vision but they differ very much in strategy, starting hypothesis and definition (frequently implicit) of `eye'. And, unfortunately,  almost none of them emphasize the invariants (intrinsic properties) of the configuration studied,  a crucial fact for us.%
\footnote{It would be stimulating to analyse and classify all this material, and select those result attached to hypergon eyes.}
\begin{figure}[h] 
\centering 
{\includegraphics[width=0.40\textwidth,height=0.65\columnwidth] 
{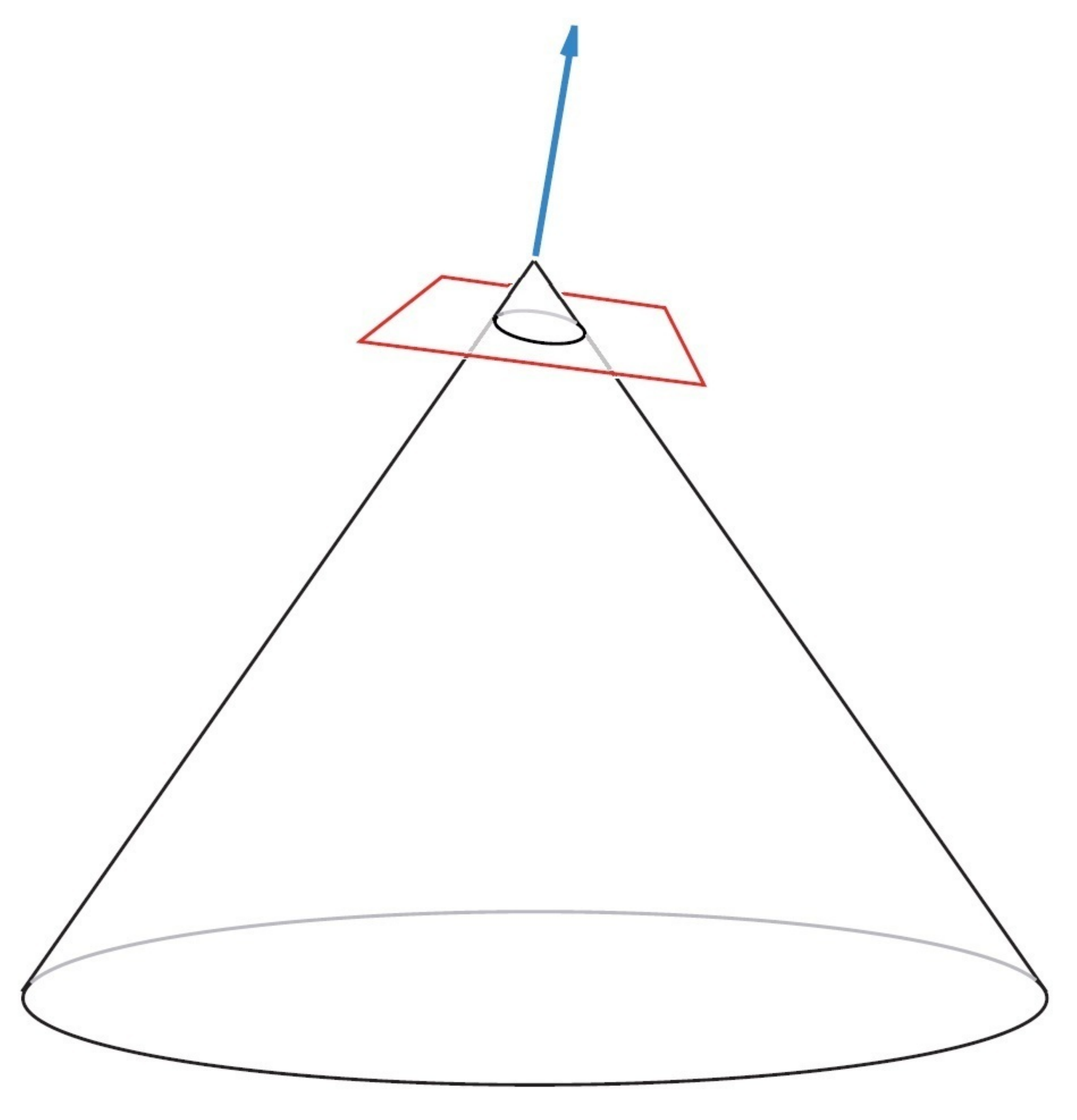}}
\caption{\small A $4\pi$-wide hypergon eye as a local device that projects the past-like cone of the event on its celestial sphere.}
\label{FgOjo}
\end{figure}  

A very interesting feature here is that {\em relativistic stereometric systems are the causal duals of positioning systems}. 
\begin{figure}[h] 
\centering 
{\includegraphics[width=0.40\textwidth,height=0.65\columnwidth] 
{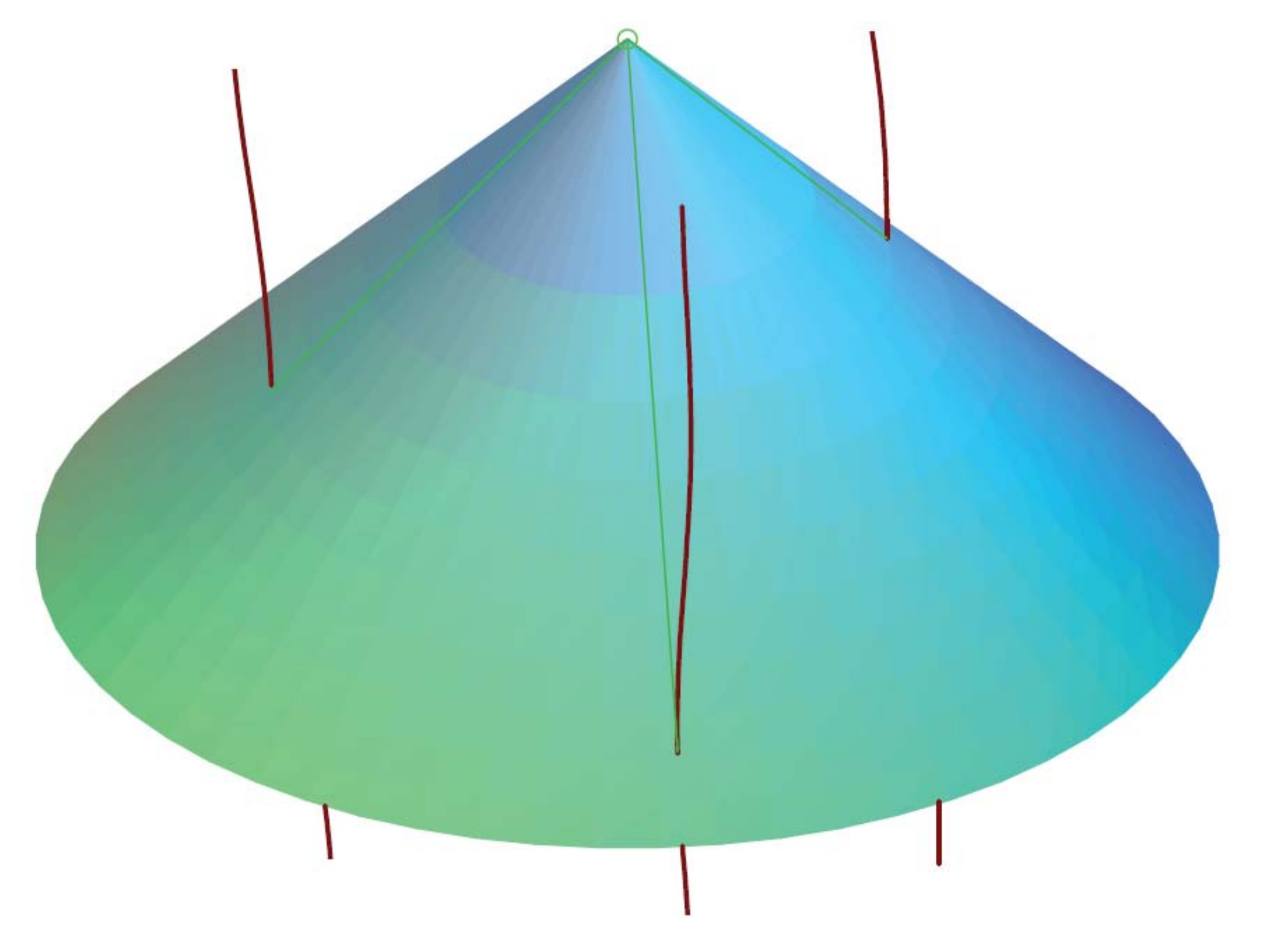}}
\caption{\small Three-dimensional representation of a relativistic positioning system.}
\label{FgRPS3d}
\end{figure}  

\begin{figure}[h] 
\centering 
{\includegraphics[width=0.40\textwidth,height=0.65\columnwidth] 
{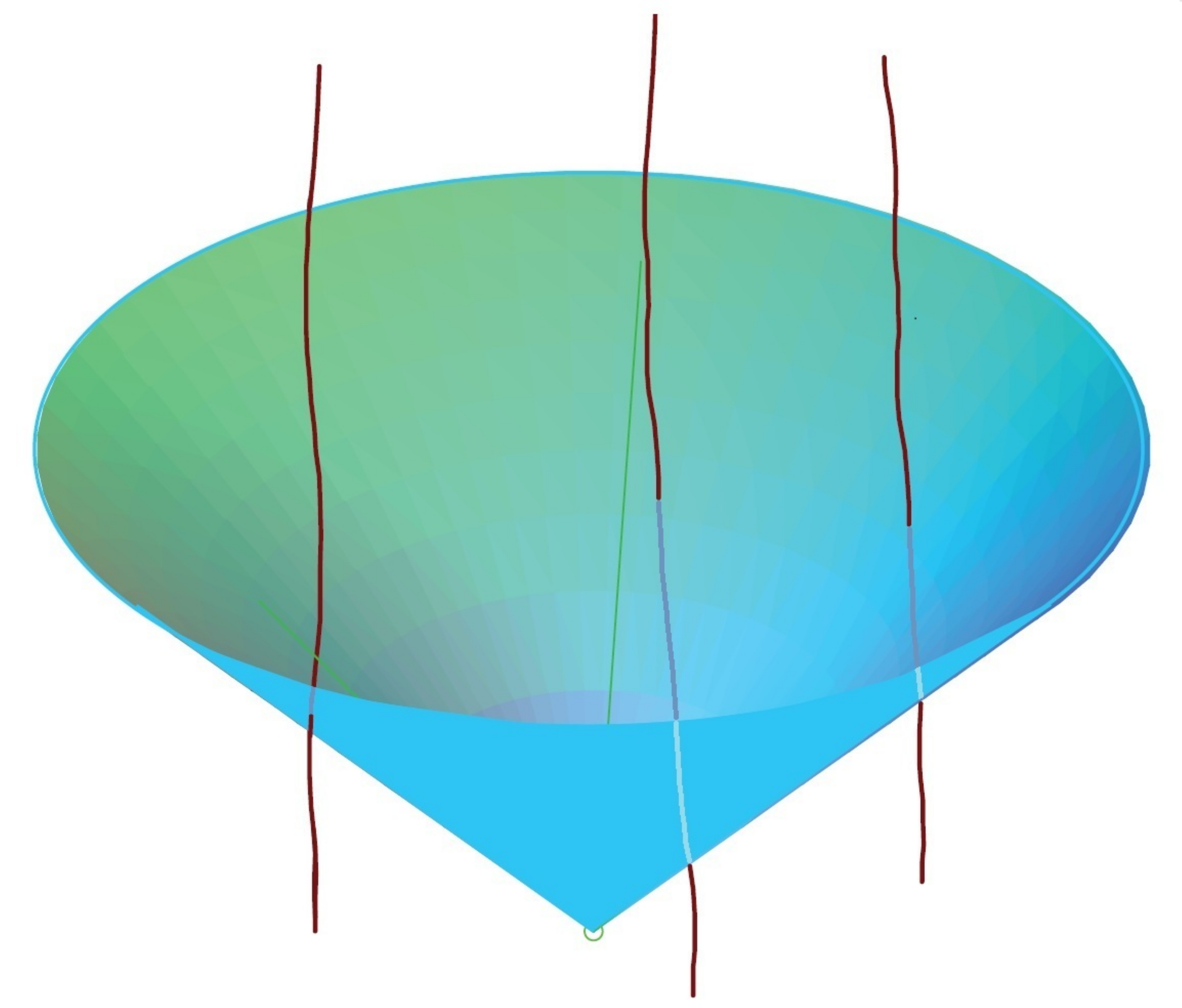}}
\caption{\small Three-dimensional representation of a relativistic stereometric system.}
\label{FgRES3d}
\end{figure}

They are causal duals operationally, relativistic positioning systems are {\em passive} for the user meanwhile relativistic stereometric systems are {\em active},%
\footnote{A relativistic stereometric system is also a {\em location system} for {\em active} users, i.e.  for those emitting an instant-identifier, as for example a clock. The times of reception, by the four observers of the stereometric system, of the signal of an instant of the user, constitute the {\em reception coordinates} of the user at that instant. Such a reception system, in addition to be active, is also not immediate: it is not a relativistic positioning system.}   
but also conceptually, as space-time objects, because they differ simply by a timelike inversion (see Figs \ref{FgRPS3d} and \ref{FgRES3d}). It is then clear that many of the properties of one of these systems may be transformed, by simple  change of time orientation, in properties of the other system.

The aim of relativistic stereometry is the obtention of the intrinsic properties of physical systems starting from their relative properties seen by four observers. 

In relativity, because the space-time objects are {\em histories},%
\footnote{Namely, they are  the histories of the spatial objects of Newtonian theory.}  
the intrinsic properties of a system related to its {\em form} must involve, besides the field of proper distances between its neighbouring points, the field of proper times of its local elements.  A main set of intrinsic properties are the {\em visual} ones, obtained by adding to the intrinsic geometric properties, the colour field of the local elements. 

Thus a first basic problem of relativistic stereometry is to obtain, of every local element of a physical system, the proper distances to its neighbouring elements and its proper colour from the corresponding relative elements observed by the four observers of the relativistic stereometric system. 
\subsubsection{First theorems in \\relativistic stereometry}
The intrinsic properties of a system being those which are observer-invariant, in relativistic stereometry we have to solve, starting from 
four observer-dependent perspectives, an inverse problem.

In order to show how it works, we shall consider the simplest  physical system in the simplest stereometric operational frame: a coloured point particle in a two-dimensional Minkowski space-time.  In spite of its easy framework, we shall see that  the solution to this stereometric inverse problem is interesting enough and constitutes a good example of epistemic relativity.
\begin{figure}[h] 
\centering 
{\includegraphics[width=0.30\textwidth,height=0.65\columnwidth] 
{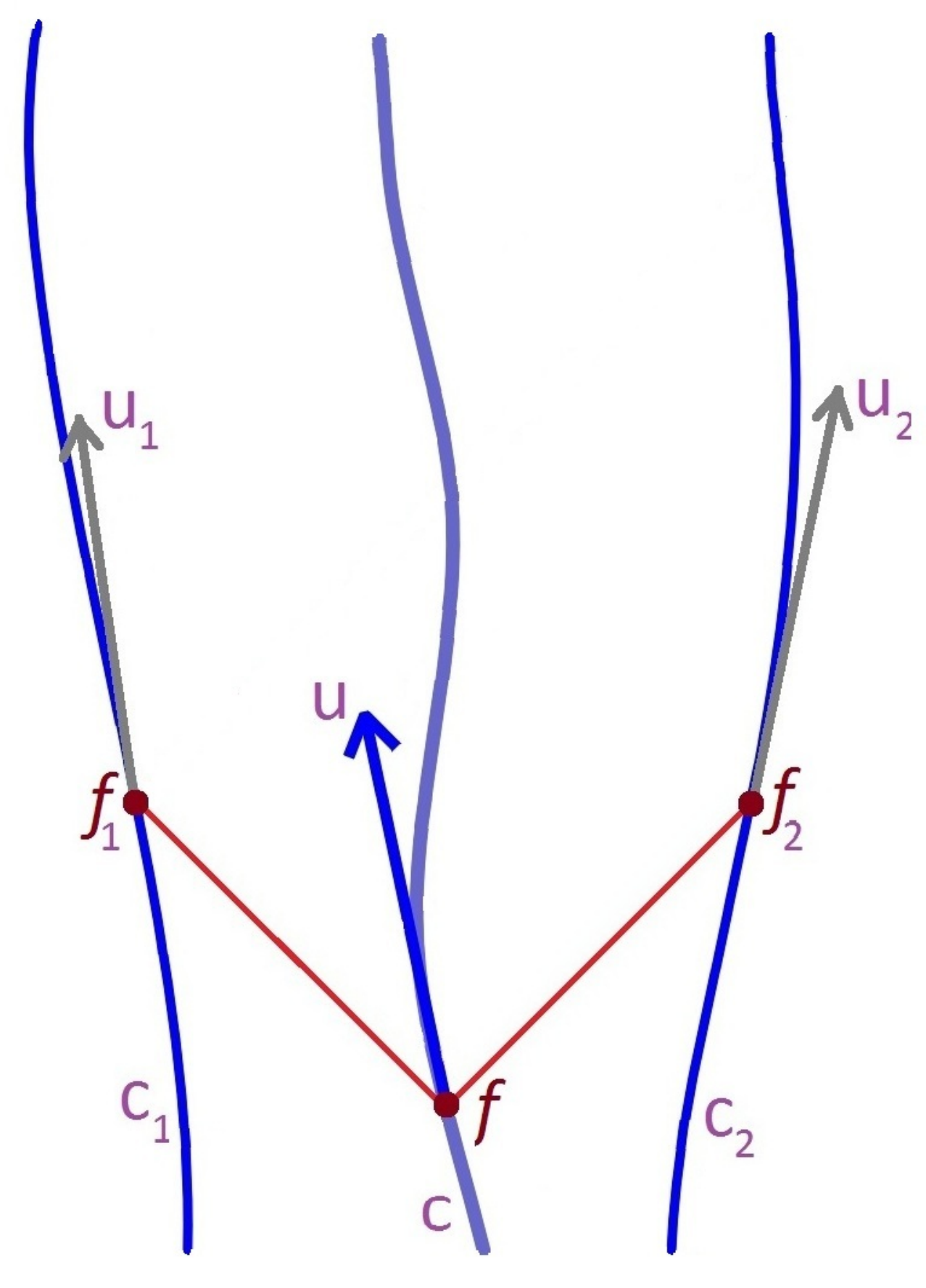}}
\caption{\small A two-dimensional relativistic stereometric system obtaining the proper frequency of a particle.}
\label{Fg2DStereo}
\end{figure}  

Thus, in Minkowski two-dimensional space-time, let $C$  be the world-line of a colored material point of a physical system, of proper frequency $f$.
Let $f_1$  and $f_2$ be the Doppler frequencies received by the relativistic stereometric system from an instant of $C$%
\footnote{There is no matter here what instant-identifier is used:  a clock associate to the point, measuring any time, non necessarily proper, a flash, or any other pertinent one.}  
and let $v_{12}$ be the relative velocity of the observers $C_1$ and $C_2$ at the instants of reception of the signals $f_1$ and $f_2$. Then we have:

{\bf Theorem 1.-} {\em In terms of the received frequencies $f_1$  and $f_2$  and of the relative velocity $v_{12}$ of the system at the reception instants, the proper frequency  $f$  of the colored point $C$ is given by:}
\begin{equation*}\label{SquaredFrequency} 
f^2 = f_1 f_2 \sqrt{ \frac{1 +v_{12}}{1 - v_{12}}} 
\end{equation*}

Observe that, if the coloured point $C$ is transported by one of the observers ($f$ $= f_i$ for some $i$ $= 1,2$), the above expression reduces to the standard one for the Doppler shift.

In addition to the proper frequency, the Doppler frequencies $f_1$  and $f_2$ also allow to work out the relative velocities of the material point:

{\bf Theorem 2.-} {\em  The relative velocities $v_1$ and $v_2$ of the material point $C$ with respect to the observers  $C_1$ and $C_2$ of the relativistic stereometric system at the instants of reception of the signals $f_1$ and $f_2$ are given by:}
\begin{equation*}\label{RelativeVelocities} 
v_1 = \frac{f_2\sqrt{1 + v_{12}} - f_1 \sqrt{1 - v_{12}}}{f_2\sqrt{1 + v_{12}} + f_1 \sqrt{1 - v_{12}}} 
\end{equation*}
\begin{equation*}
v_2 = \frac{f_1\sqrt{1 + v_{12}} - f_2 \sqrt{1 - v_{12}}}{f_1\sqrt{1 + v_{12}} + f_2 \sqrt{1 - v_{12}}} 
\end{equation*}

Observe that the results in both theorems depend not only of the measured Doppler frequencies, but also of the relative velocity $v_{12}$ of the observes of the stereometric system at the instants of reception of these frequencies, a quantity that seems not obvious how to be measured. The question is thus: are these theorems epistemic ?

In fact, they {\em are not} epistemic. Moreover: by themselves they {\em cannot be} epistemic. The simple reason is that, without additional specifications, these two theorems do not fulfill any of the above  three conditions characterizing epistemic relativity. 

\begin{figure}[h] 
\centering 
{\includegraphics[width=0.30\textwidth,height=0.65\columnwidth] 
{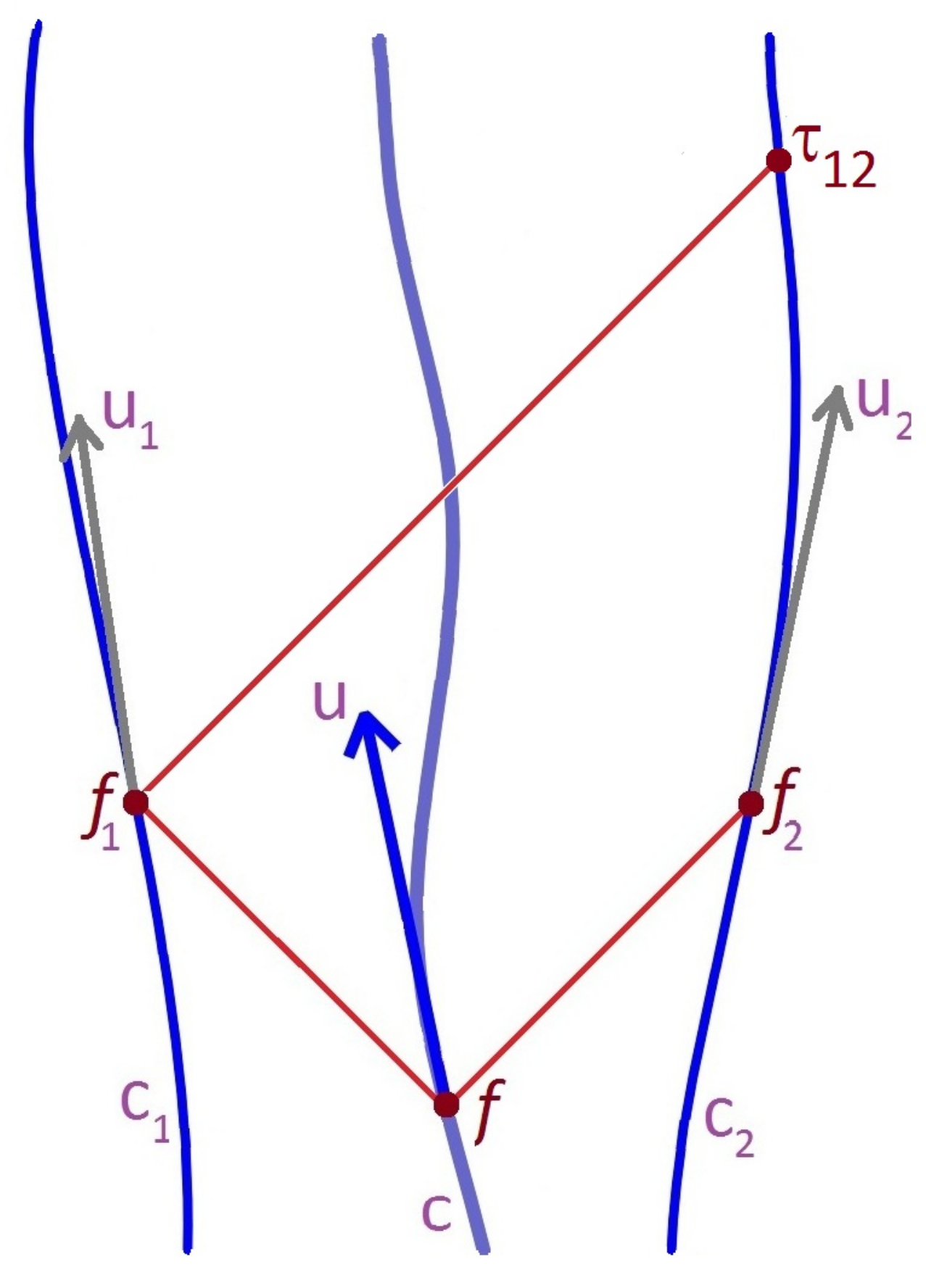}}
\caption{\small Here, $C_2$ has been chosen as the physicist of the epistemic problem and, from the instant $\tau_{12}$  on, it may know all the quantities of the problem in question.}
\label{Fg2DStereoTaus}
\end{figure}

To fulfill these conditions, we must complete the above results with the information about:
\begin{itemize}
	\item  what physicist  we have choose to make the experiment,
	\item when and where it%
\footnote{The world {\em physicist} here denotes any person or device able to receive the pertinent information from the relativistic  stereometric system, to record and to analyze it and to perform the computations needed for the problem in question. For short, we shall refer to this physicist as {\em it}. } 
 is able to be informed of the quantities needed to answer the problem. 
	\item  how can it know or measure these quantities.	
\end{itemize}
 This information is chosen here as follows:
\begin{itemize}
	\item[*] the simplest choice of physicist is to take it as one of the observers of the relativistic stereometric system, say $C_2$, as shown in Fig  \ref{Fg2DStereoTaus}, 
	\item[*] then it will be able to be informed of all the quantities needed to answer our problem at the instant $\tau_{12}$ of reception of the pertinent information coming from the observer $C_1$, 
	\item[*] at that instant $\tau_{12}$ it is informed of the quantity $f_1$, it already has the quantity $f_2$, mesured and recorded by it, and it may know the quantity  $v_{12}$  by computation. 
\end{itemize}

The computation of the relative velocity $v_{12}$ may be made, for example,  from the knowledge of the worldlines of the observers of the relativistic stereometric system.  In the case where these observers are geodesic, its expression is very simple. If, for short,  we call an epistemic theorem an {\em epistem}, from the above two theorems we have: 

{\bf Epistem 0.-} {\em In terms of the frequency of the proper time of the observer $C_1$ with respect to the proper time of the observer $C_2$, $\nu_{12}$, the relative velocity $v_{12}$ between the observers of a geodesic relativistic stereometric system is given by:}   
\begin{equation*}\label{RelativeVelocity} 
v_{12} = \frac{1 - \nu_{12}^2}{1 + \nu_{12}^2} \; . 
\end{equation*}

Now, for this geodesic case, the above two theorems become respectively:

{\bf Epistem 1.-} {\em In terms of the received frequencies $f_1$ and $f_2$ of a coloured point $C$ and of the frequency of proper times $\nu_{12}$ of a geodesic stereometric system, the proper frequency $f$ of a `coloured'  point $C$ is given by:}
\begin{equation*}\label{episten1} 
f^2 =  \frac{f_1 f_2}{ \nu_{12}}   \; .
\end{equation*}

{\bf Epistem 2.-} {\em The relative velocities $v_1$ and $v_2$ of the material point $C$ with respect to the observers $C_1$ and $C_2$ of a geodesic stereometric system at the instants of reception of the signals $f_1$ and $f_2$ are given by:}  
\begin{equation*}\label{epistem2} 
v_1 = \frac{f_2 - f_1 \nu_{12}}{f_2 + f_1  \nu_{12}} \, ,
\end{equation*}
\begin{equation*}
v_2 = \frac{f_1 - f_2 \nu_{12}}{f_1 + f_2\nu_{12}} \, . 
\end{equation*}

In spite of the very simple context in which they are obtained, these resuts are interesting because they show the essentials of epistemic relativity.  Of course, because the celestial sphere of an observer in a bidimensional space-time reduces to two opposite points, the problem of determining the proper distances to its neighbouring elements of a material point cannot be considered in this dimension. The need of extending this work to three or four dimensions is evident.
\subsection{What about the mathematics of relativity?}
The development of Riemannian geometry at the beginning of the 20th century, the already existing classical theory of deformations and (the belief in) the local character of the fundamental physical laws are at the basis of the mathematical frame of the general theory of relativity, namely the Lorentzian differential geometry. 
\subsubsection{Insufficiency of differential \\geometry in relativity}
But, in its present form, in front of these historical and conceptual justifications, there exist practical (epistemic!) insufficiencies of differential geometry in relativity. Initially, differential geometry appears appropriate for the description of local curved regions, but classical fields are of infinite range, so that little local perturbations of a physical system do not remain confined, but spread indefinitely. This propagation of local little perturbations cannot be neglegted, not only because of it physically meaningful character, but because it do not take place subtly, but at the velocity of light, an antroposcopic%
\footnote{Visible to the naked eye and important enough for human activities.} 
velocity.%
\footnote{Think, simply, in the effect of strucking a match in the darkness.}  

These general facts do not dimisish the unavoidable character of differential geometry, its inevitability in the formulation of relativity theory, but show its {\em insufficiency}. More particularly, in the estudy of positioning or stereometric systems, or simply in the study of any epistemic problem of finite extension, this insufficiency is dramatic. We are all suffering of this situation%
\footnote{Besides this insufficiency, one could add the absolute lack of covariant methods of perturbations and approximations. After beautiful discourses about the importance of the role of the geometrization of physics by relativity, the most simple approximate calculation or deformation of a metric is made, without embarrassment, with non covariant analytical methods devoid of geometrical meanings. It is clear that the usual mathematical methods in relativity are not well adapted to relativity. But this subject will not be considered here.} . 
\subsubsection{Finite-differential geometry}
I think that it is an urgent task in relativity, for all of us, to try to construct a {\em  finite-differential geometry.}

The purpose of finite-differential geometry is to introduce  interchangeable finite versions of the basic ingredients of differential geometry, namely:
\vspace{-2.5mm}
\begin{itemize}
	\item[*] metric $g$,
\vspace{-2.5mm}
	\item[*] connection $\Gamma$, 
\vspace{-2.5mm}
	\item[*] curvature  $Riem$.
\end{itemize}
\vspace{-1mm}

The distance function  $D(x,y)$, or its half-square $\Omega(x,y)$,  the Synge's  {\em world-function}, 
\begin{equation*}\label{worldfunction} 
\Omega(x,y) = \frac{1}{2} D(x,y)^2  \; ,
\end{equation*}
are already finite versions of the metric  g. As it is known:
\begin{equation*}\label{worldfunctioninteg} 
\Omega(x,y) = \frac{1}{2}\int_0^1 g( \frac{d\gamma}{d\lambda},\frac{d\gamma}{d\lambda})d\lambda   \; ,
\end{equation*}
$\gamma(\lambda)$ being the geodesic joining $x$ and $y$, and their fundamental equations are:
\begin{equation*}\label{worldfunctionfundeqs} 
g^{\alpha\beta}\partial_\alpha\Omega \partial_\beta\Omega = 2\Omega \; , 
g^{ab}\partial_a\Omega \partial_b\Omega = 2\Omega \; ,
\end{equation*}
where Greek and Latin indices are related to the first and second arguments of $\Omega(x,y)$ respectively.

Distance spaces are well known, but their link with differential geometry has not been yet sufficiently explored.

Let us think, in a given space-time, on a positioning system complemented with a number of additional clocks. This over-determined system will generate an over-determined set of data able to select a distance function with some uncertainty. Well, in spite of its interest, this problem is open for space-time distances.

The first problem to be solved for any proposed%
\footnote{Or experimentally obtained with some uncertainty.}   
distance function is if it is really the (geodesic) distance function of a metric. The constraints for this to be the case will constitute an important tool to improve uncertainties and to delimit parameter values.  

I solved this problem some years ago. In order to express its solution, it is convenient to introduce some algebraic functions of the first and second partial derivatives of the symmetric bifunction $D(x,y)$ $ \equiv D$  proposed as distance function. Remember that at this level we have no metric at all, and that all subscripts in $D$ denote partial derivatives. Define $V_{abc}^\alpha$ as the following function of second order derivatives of $D$:
\begin{equation*}\label{Vabcalfa} 
V_{\ell mn}^\alpha \equiv \epsilon^{\alpha\lambda\mu\nu}D_{\ell\lambda} D_{m\mu} D_{n\nu}
  \; ,
\end{equation*}
$V^{a\alpha}$ as the following combination of first and second ones:
\begin{equation*}\label{Vaalfa} 
V^{a\alpha} \equiv   \epsilon^{a\ell mn}  \epsilon^{\alpha\lambda\mu\nu}D_\ell D_\lambda  D_{m\mu} D_{n\nu}   \; .
\end{equation*}
and $V^\alpha$ as the quantity:
\begin{equation*}\label{Vabcalfa} 
V^\alpha \equiv V_{\ell mn}^\alpha x^\ell y^m z^n    \; ,
\end{equation*}
where $x^\ell$, $y^m$, $z^n$ are arbitrary independent directions. Introduce the two scalars
\begin{equation*}\label{fiandpsi} 
\Phi \equiv D_\lambda V^\lambda  \; ,  \Psi \equiv  \epsilon^{r\ell mn} V_{\ell mn}^\rho D_r D_\rho 
  \; ,
\end{equation*}
and form the two quantities:
\begin{equation*}\label{Dsupers} 
D^\alpha \equiv \frac{V^\alpha}{\Phi}   \; ,  \;
D^{a\alpha} \equiv  3 \frac{V^{a\alpha}}{\Psi} \; .
\end{equation*}
Then, we have:

{\bf Theorem 3.-} (structure theorem for distance functions) {\em The necessary and	sufficient condition for a distance function $D(x,y)$  to be the geodesic distance function of a metric, is that its derivatives verify the identity:}
\begin{eqnarray*}\label{distfunctconstraint} 
D_{abc\rho} D^\rho + D_{(ab|\rho|}D_{c)m\sigma} D^{m\rho} D^\sigma \\
- D_{(ab|\rho|} D_{c)} D_{mn\sigma} D^{m\rho} D^n D^\sigma = 0 \; ,
\end{eqnarray*}
{\em where the subscripts denote partial derivatives and $D^a$ and $D^{a\alpha}$ are the quantities just defined.}

Note that, in any of the above expressions Latin and Greek indices can be exchanged because the symmetric character of the distant function proposed. Another point to be noted, very little known, is that  a function may be independent of some parameters meanwhile its algebraic expression depends unavoidably of them, if these parameters are not scalars. It is here the case of $D^\alpha$ whose expression, from the definition of $V^\alpha$, depends of the arbitrary vectors $x^\ell$, $y^m$, $z^n$ but that, as a function, it does not dependent of them: the partial derivatives of $D^\alpha$ with respect to any of these vectors vanishes.

Once we know that a proposed distance function is truly%
\footnote{At the admissible uncertainties.}   
a geodesic distant function of some metric, the second problem to be solved is to obtain that metric. If, for example,  as it may correspond to a natural experimental protocol, the proposed distance function $D(x,y)$ has been obtained from two local groups of points separated by non local distances, very probably the standard method of obtaining the metric by taking the limit when a point of one local group, say $y$,  reaches a point of the other, say $x$, may have no sense, neither physical nor mathematical. For this reason, one needs to obtain the metric by means of a {\em finite} method. I did that some years ago, and the result is:

{\bf Theorem 4.-} (metric of a  distance function) {\em  In terms of the derivatives of the distance function $D$, the contravariant components  $g^{\alpha\beta}$ of the metric solution at the point  $x$  are given by:}
\begin{equation*}\label{metricfromdistance} 
g^{\alpha\beta} = D^\alpha D^\beta + D^{a\alpha} D^{b\beta} D_{ab\gamma}D^\gamma \; .
\end{equation*}

Note that the right hand side of this equation is a combination of a symmetric bifunction $D(x,y)$ and its partial derivatives, which, in general is other bifunction, meanwhile the left hand side is a function of the sole variable $x$. There is no contradiction: there are the conditions of theorem 3 that guarantee the downfall of the variable $y$. 

This expression is very well adapted for the computation from an approximate distance function by means of finite difference methods.				
					     
The finite analog of a connection remains a completely open problem. Perhaps this problem is avoidable, but it is not avoidable the quest for a finite version of curvature, because curvature is directly related to the energetic content of physical fields. With my friend Albert Tarantola (1949-2009), in the lustrum 2001-2005, we associated, to every four elements of a space-time, a finite object that seems to be a finite definition of curvature but, unfortunately, we were not able to prove that it is so. 

The above results remain, for our needs, elemental. We must still develop them, plan a phenomenology of distance functions, and learn to be able to ask it the same finite questions that we are asking to a geometric structure.

\vspace{6mm}
{\bf Acknowledgments}
This work has been supported by the Spanish ministries of ``Ciencia e Innovaci\'on  and ``Econom\'{\i}a y Competitividad'', MICINN-FEDER projects FIS2009-07705 and FIS2012-33582.

\end{document}